\documentstyle[prl,aps,psfig,preprint]{revtex} 

\newcommand{\bra}[1]{\overline{#1}}

\begin{document}
\draft
\title{Localization induced by noise and non linearity}
\author{Ph. Blanchard, $^{(1)}$, M. Pasquini$^{(2)}$ and 
M. Serva$^{(2,3)}$}
\address{$^{(1)}$Fakult\"at f\"ur Physik, 
Universit\"at Bielefeld, D-33615 Bielefeld, Germany}
\address{$^{(2)}$Istituto Nazionale Fisica della Materia,
Universit\`a dell'Aquila, I-67010 Coppito, L'Aquila, Italy}
\address{$^{(3)}$Dipartimento di Matematica, Universit\`a 
dell'Aquila, I-67010 Coppito, L'Aquila, Italy}

\date{\today}

\maketitle

\begin{abstract}
We introduce a model for a two configurations system, 
and we study the transition from quantum to classical behaviour.
We first consider the effect of 
the interaction with the environment as an external noise 
and we show that it produces decoherence and suppression of tunnelling.
These features are widely accepted as definition of {\it classicality},
while we believe that {\it classicality} implies that
quantum delocalized states spontaneously evolve into localized ones.
We than show that this evolution take place only when both
noise and non linearity in the equations are present.

\end{abstract}


\narrowtext

In this paper we focus our attention on a very old question: 
how and why macroscopic objects behave classically.
If we disregard any answer which invokes a non-quantum
or quantum-modified status for macroscopic objects
\cite{Gisin,GRW}, 
we remain with the widely accepted point of view that
environment induces decoherence.
This fact, which has been shown to hold in a huge number of models
(see for example \cite{Zurek,ChKi,ChLe,book,GDJH}), 
can be stated as follows: 
the density matrix corresponding to a superposition 
of eigen-states of a macroscopic variable becomes very soon 
almost diagonal.
Therefore, the superposition is {\it for all practical purposes}
indistinguishable from the statistical mixture,
whose probabilities are given by the diagonal elements
\cite{Bell}.

Our point is that this is not a complete answer, since it has 
only a statistical meaning, it remains to be explained why
individual macroscopic objects are always found in a localized state.
If we want such an explanation we are left with only a possibility:
to show that any superposition quantum state spontaneously 
evolves in a eigen-state of the macroscopic variable.

We study the problem by considering a system
with two minimal energy configuration positions.
Systems of this kind are, at low temperature, 
classically localized in one of the two minima.
On the contrary, quantum behaviour allows for 
delocalized states and coherent tunnelling.
Real objects of this type are, for example, superconducting 
rings with a {\it weak} junction crossed by a magnetic flux
or a class of tetraedrical molecules. 

We start considering {\it ab initio} a two configurations model. 
The isolated system tunnels coherently and periodically. 
Then we consider the effect of a time dependent perturbation
(interaction with the environment). Now there is decoherence
and, for strong noise, suppression of tunnelling 
(see also \cite{BBCdAS,Berry,CFPSV}). 

Our point, as explained before, is that decoherence and 
suppression of tunnelling which implies stability of localization 
are not enough for {\it classicality}. 
{\it Classicality} needs spontaneous evolution
into one of the two possible localized states.

Therefore, in a next step we consider the same model, but we assume 
a non linear evolution. In absence of perturbation, 
it turns out that the system behaves qualitatively as in the 
linear unperturbed case. But, this time, if interaction 
with the environment is introduced,
one has that there is spontaneous evolution into
one of the two localized states.
The conclusion is that both non linearity and noise are fundamental 
ingredients for {\it classicality}.

\bigskip
The density matrix 
can be written in a convenient form using 
the vector ${\bf x} =(x,y,z)$ defined by 

\begin{equation}
\rho= \frac{1}{2}
\left( 
\begin{array}{cc}
1+z   & x-iy \\
x+iy  & 1-z 
\end{array} 
\right)
\label{dm}
\end{equation}
The trace of this density operator equals 1 (normalization) 
and the determinant vanishes for a pure state.
This last requirement implies $|{\bf x}|=1$ 
(a pure state is represented by a point on the unitary sphere).

The third component $z$ of this vector encodes the informations
about the localization ($z = \pm1$ means that the
system is exactly localized in one of the two states) while
$x$ and $y$ encode the informations about coherence
(when they vanish there is complete decoherence).

The above considerations show that the quantum evolution
$\dot{\rho} = - i [ H , \rho ]$ corresponds to
the motion of a point on a unitary sphere. When the point
is in one of the two poles the system is localized.

The hamiltonian which describes an isolated two configurations 
system allowed for tunnelling is typically 
\begin{equation}
H = {\alpha} {\sigma}_x 
\label{hf}
\end{equation}
where $\sigma_x$ is the Pauli matrix.
This hamiltonian produces coherent 
quantum tunnelling of period
$\frac{\pi}{\alpha}$ between the two
{\it macroscopic} configurations of the system
(in mathematical terms, the two eigen-states
of the Pauli matrix $\sigma_z$).

The corresponding equations for ${\bf x}$ are
\begin{equation}
\begin{array}{lll}
\dot{x} & =  0 \\ 
\dot{y} & =  - 2 \alpha z   \\
\dot{z} & =    2 \alpha y 
\end{array}
\label{rf}
\end{equation}

It is clear that (\ref{hf}) induces a rotation with constant angular 
velocity around the $x$ axis. The value of the first component
of the vector remains unchanged. Therefore, (\ref{hf})
never allows for localization or decoherence.

Assume now that the interaction with the environment is described 
by

\begin{equation}
H = {\alpha} {\sigma}_x \ + \ {\epsilon}(t) {\sigma}_z 
\label{he}
\end{equation}
where $\sigma_x$ and $\sigma_z$ are Pauli matrices
and $ \epsilon(t)$ is a a given realization
of a stochastic process.
The presence of a non vanishing $\epsilon(t)$
means that we introduce an interaction
with the environment which
randomly breaks in time the energy symmetry of the
two configurations.

We find that the vector ${\bf x} $ satisfies

\begin{equation}
\begin{array}{lll}
\dot{x} & =  - 2 \epsilon(t) y  \\ 
\dot{y} & =  - 2 \epsilon(t) x \ - \ 2 \alpha z   \\
\dot{z} & =  + 2 \alpha y  
\end{array}
\label{re}
\end{equation}
which is the superposition of two rotations,
the first around the $x$ axis with constant angular 
velocity $2\alpha$, the second around the $z$ axis with 
time dependent angular velocity $2\epsilon(t)$.

Assume $\epsilon(t) = \beta \eta(t)$ where
$\eta(t)$ is a given realization of a white noise (i.e. $w(t)\equiv
\int_0^t \eta (s) ds$ is a brownian motion).
In Stratanovitch notation the only thing we have to do
is to substitute $\epsilon(t)$ with $ \beta \eta(t)$
in the above equations. Nevertheless, it is much more practical
to use Ito notation for which the above equations (\ref{re})
rewrite as
\begin{equation}
\begin{array}{lll}
dx & =  - 2 \beta^2 x dt \ - \ 2 \beta  y dw \\ 
dy & =  - 2 \beta^2 y dt \ - \ 2 \alpha z dt 
\ + \ 2 \beta  x  dw  \\
dz & =  + 2 \alpha y dt
\end{array}
\label{rito}
\end{equation}

From (\ref{rito}), taking the expectation value, we immediately 
obtain
\begin{equation}
\begin{array}{lll}
\dot{\bra{x}} & =  - 2 \beta^2  \bra{x} \\
\dot{\bra{y}} & =  - 2 \beta^2  \bra{y} \ 
- \ 2 \alpha \bra{z} \\ 
\dot{\bra{z}} & =  + 2 \alpha  \bra{y}  
\end{array}
\label{ritom}
\end{equation}

Since the density matrix depends linearly on ${\bf x}$,
the above equation describes the evolution of the averaged
density matrix $\bra{\rho}$. This matrix describes the mixture 
of all the states associated to the 
ensemble of all the realizations of the noise
in the  hamiltonian (\ref{he}).

Equation (\ref{ritom}) can be easily solved.
When $\beta^2<2 |\alpha|$ the solution is:
\begin{equation}
\begin{array}{lll}
\bra{x(t)} & =  e^{-2\beta^2 t} x(0)  \\
\bra{y(t)} & =  e^{-\beta^2 t} [ y(0)  \cos (\omega t) +
c_1 \sin (\omega t)]  \\
\bra{z(t)} & =  e^{-\beta^2 t} [ z(0)  \cos (\omega t) + 
c_2 \sin (\omega t)] 
\end{array}
\label{sm1}
\end{equation}
where $\omega= \sqrt{|\beta^4 - 4\alpha^2|}$, $c_1 = 
\frac{ - \beta^2 y(0)  - 2 \alpha z(0) }{ \omega }$
and $c_2 = 
\frac{ \beta^2 z(0) + 2\alpha y(0) }{ \omega }$.
Looking at the third component
$z(t)$ one realizes that in this region 
a quantum coherent behavior partially survives to the noise
for a certain time.

When $ \beta^2 > 2 |\alpha| $ the solution is purely exponential 
and  is given by
\begin{equation}
\begin{array}{lll}
\bra{x(t)} & =  e^{-2\beta^2 t} x(0)  \\
\bra{y(t)} & =  e^{-\beta^2 t} [ y(0)  \cosh (\omega t) +
c_1 \sinh (\omega t)]  \\
\bra{z(t)} & =  e^{-\beta^2 t} [ z(0)  \cosh (\omega t) +
c_2 \sinh (\omega t)]
\end{array}
\label{sm2}
\end{equation}

In this strong noise region the coherent behaviour is
completely destroyed since
the localization probability relaxes
exponentially.
The interesting fact is that for large $\beta$, $ \bra{z(t)}$
relaxes with an exponent which vanishes as $\frac{2 
\alpha}{\beta^2}$.
This fact means that the time necessary for delocalization 
becomes infinite in the limit of extremely strong noise.

Summarizing, we have damped oscillations for a weak interaction 
with the environment ($\beta^2 < 2 |\alpha|$) and 
incoherent relaxation for strong interaction ($\beta^2 > 2 |\alpha|$).
Furthermore, in the limit of very strong noise a localized configuration
turns out to be almost stable.

Since in both cases $ \bra{{\bf x}} \to 0$ exponentially, we have 
that 
\begin{equation}
\bra{\rho} \to \left(\matrix{{1\over2} & 0 \cr 0 & {1\over2}\cr} 
\right)
\label{dml}
\end{equation}
which says that in any case 
we have almost complete decoherence (no diagonal terms)
at large times, the non-diagonal elements being damped whereas the 
diagonal ones carry the probabilities. 

From (\ref{dml}) 
we could conclude that an individual system is in one of the two
configuration positions. This is not true, in fact,
from the stochastic equations (\ref{rito}) one can derive the set
of linear equations
\begin{equation}
\begin{array}{lll}
\dot{ \bra{z^2}} & =  4 \alpha \bra{zy}   \\
\dot{ \bra{y^2}} & =  - 4 \alpha \bra{zy} -  4 \beta^2 \bra{y^2} 
\ + \ 4 \beta^2 \bra{x^2} \\
\dot{ \bra{yz}} &
=  - 2 \beta^2 \bra{zy} \ + \ 2 \alpha \bra{y^2}
- \ 2 \alpha \bra{z^2} 
\end{array}
\label{2m}
\end{equation}
Using the condition $x^2 + y^2 + z^2 = 1$, one can replace 
$\bra{x^2}$ with $1 - \bra{y^2} - \bra{z^2} $
in the second of the above equations. 
For any $\beta \neq 0$ one finds that the
solution converges to the stationary solution 
$\bra{x^2} = \bra{y^2} 
= \bra{z^2}  =\frac{1}{3}$,
$\bra{z y} = 0$.
This means that the typical state is not localized 
(localization for all states would imply 
$\bra{z^2}  =1$).
A new ingredient is thus necessary to produce localization.

\bigskip

The effect of the interaction with the environment 
can be taken into account also considering a non linear term
in the differential equations that describe the evolution of
the density matrix. In our case, we can assume that 
$\alpha$ depends on the state ($\alpha = \alpha(x,y,z)$).
For the sake of simplicity assume, as usual, that this
dependence reduces to a dependence on the 
modulus square of the wave function. In our language it means that
$\alpha$ depends only on $z$. 
There are many reasons for this choice, that we 
will discuss in a forthcoming paper \cite{BDAPS}.

Also assume that:

a) $\alpha(z)$ is an even function of $z$: $\alpha(z)=\alpha(-z)$;

b) $\alpha(z)$ is zero at the poles: $\alpha(+1)=\alpha(-1)=0$, 
positive otherwise and sufficiently smooth around the poles.

\noindent
The requirement b) implies that 
a localized state remains, in fact, localized.
Nevertheless, in absence of noise, 
the system is not able to spontaneously localize.
In this case, in fact, equation (\ref{rf}) still holds, with the only
difference that $\alpha$ depends on $z$.

Notice that the first component $x$ of the vector ${\bf x}$ 
remains constant during the motion. This means that there is 
no decoherence, and, furthermore, that the system can never 
reach one of the poles.
Indeed, the motion remains periodic, since we are on a two 
dimensional variety.
The only exception is when the system is initially on the meridian $x=0$,
in which case it moves along this meridian
toward one of the poles, producing localization,
but this ensemble of initial conditions has probability zero.

In conclusion, non linear differential equations, 
like the equation modelizing the interaction with the environment,
are not able to reproduce a classical behaviour 
in the more restricted sense of induced localization.
Nevertheless, this goal can be achieved if one takes into account
the simultaneous presence of these two different mechanisms.

\bigskip

Let us consider again the hamiltonian (\ref{he}) with
$\alpha=\alpha(z)$; the previous requirements 
on $\alpha(z)$ are also assumed. 
The differential equations (\ref{rito}),
which remain unchanged, have two fixed points
corresponding to the poles. What we can show
is that, independently on the initial conditions, one has 
$z(t) \to\pm 1$. This convergence is not due to
the attractiveness of the poles but to the fact
that the motion along the meridians becomes more and more slow 
approaching the poles. 
In other words, the system stays for long time intervals around them,
the distribution of this time intervals having sufficiently long
tails to guarantee that, at the end the system will spend 
almost all its time around a pole 
(it will be almost surely in a pole).

We can sketch how to show that the system localizes for 
large times; a rigorous proof being beyond the scope of this paper.
The requirements on $\alpha(z)$ imply that
it can be approximated around the poles by
\begin{equation}
\alpha(z) = \alpha_0 (1-z^2) +\dots
\label{appr}
\end{equation}

We now prove that the limiting (steady) distribution is concentrated 
on the poles. Using equations (\ref{rito}) and taking the average, we 
have

$$
{{\rm d \ }\over{{\rm d} t}}
\left[ \bra{\theta(z -z_c) y} \right] = 
- 2 \beta^2 \bra{\theta(z-z_c) y} +
$$
\begin{equation}
- 2 \bra{\alpha(z)\theta(z -z_c)  z} 
+ 2 \bra{\alpha(z) \delta(z-z_c)y^2} 
\label{diff1}
\end{equation}
where $0\le z_c \le 1$, and $\theta(\cdot)$ is the step function.

We can safely assume that any initial distribution on the surface of
the unitary sphere will evolve, after a transient time, 
toward a stable distribution. For this stable distribution,
the above time derivative of
${{\rm d \ }\over{{\rm d} t}} \bra{\theta(z -z_c) y}$ vanishes.

Furthermore, from equations (\ref{rito}) it also follows that
\begin{equation}
{{\rm d \ }\over{{\rm d} t}}
\left[ \bra{\theta(z -z_c)} \right] = 2 \bra{\alpha(z) \delta(z-z_c)y} 
\label{diff2}
\end{equation}
The above equation implies for the steady distribution
$\bra{\delta(z-z_c)y}=0$, which, in turn, implies
$\bra{\theta(z-z_c)y} =0$.

This last equality, inserted in (\ref{diff1})
gives for the stationary distribution
\begin{equation}
\bra{\alpha(z)\theta(z -z_c)  z} = 
\alpha(z_c)\bra{ \delta(z-z_c)y^2} 
\label{meq}
\end{equation}
which leads to
\begin{equation}
\lim_{z_c \to 1}
\frac{\bra{\alpha(z)\theta(z -z_c)z}}{\alpha(z_c)\bra{ 
\delta(z-z_c)y^2}}
=1 
\label{meql}
\end{equation}

It is intuitive, and it
can be also easily shown (again from (\ref{rito})) 
that the relative difference between $\bra{ \delta(z-z_c)y^2}$ 
and $ \bra{ \delta(z-z_c)x^2}$
vanishes  in the limit $z \to z_c$,
in fact the random rotation around the $z$ axis becomes 
very fast compared with the rotation around the $x$ axis.
Therefore, since $x^2+y^2+z^2=1$, we can replace
 $\bra{ \delta(z-z_c)y^2}$ with
$\bra{ \delta(z-z_c)\frac{1-z^2}{2}}$
in equation (\ref{meql}), which we rewrite as 

\begin{equation}
\lim_{z_c \to 1}
\frac{\int_{z_c}^{1}(1-z^2)z\rho(z)dz}{\frac{(1-z_c^2)^2}{2}\rho(z_c)
}
=1 
\label{meql2}
\end{equation}
where we have used the expansion of $\alpha(z)\simeq \alpha_0 (1-z^2)$
around the north pole and we have introduced explicitly
the steady state probability $\rho(z)$ 
that the system is at a quote $z$.

Assuming that at $\rho(z) \simeq \frac{1}{(1-z_c)^\gamma}$
around the north pole with $\gamma < 1$, we find that the above 
equation
gives
\begin{equation}
\frac{1}{2-\gamma} =1
\label{gamma}
\end{equation}
which cannot be satisfied for any normalizable distribution.

Finally, since (\ref{meq}) is satisfied for a 
distribution which is concentrated in 
the north pole we find that this is the only possible stable distribution.
Repeating the argument for $-1\le z_c \le 0$, and taking into 
account the symmetry of the system (\ref{rito}) with respect to the
changes $y \to -y$ and $z \to -z$, one can conclude that 
the steady distribution is made of two Dirac's delta distributions
centered on the poles, 
with equal weights.

Notice that this result implies again (\ref{dml}). 
Therefore, for what concerns decoherence
nothing is changed, the important difference being that now 
$\bra{\rho^2} = \bra{\rho}$ for large times.

In fig. 1, where  $z(t)$ is plotted starting from a numerical
solution of the differential equations (\ref{rito})
for $\alpha_0=1$ and $\beta=7$, 
one sees clearly that the system spends a large part of the time 
in very narrow regions around the poles.

The time necessary to make a transition between the poles
turns out to be negligible with respect to
the long periods in which $z\simeq \pm 1$, and therefore,
if one perform an observation of the system,
it is almost sure to find it in a localized configuration.
A useful quantity in order to show this fact 
is the time average of $z(t)^2$, defined by:
\begin{equation}
L(t) = \frac{1}{t}\int_0^t [z(s)]^2 ds
\label{z2}
\end{equation}
This average tends to grow in time since 
the periods when $z(t)^2$ is substantially different from $1$
become soon negligible,
as shown by fig. 2, where $L(t)$ is plotted for the same noise realization 
as fig. 1.

In conclusion, 
we have presented a model where 
the coupling with the environment is represented both by a noisy
potential and a non linear correction to the differential equations.
This second effect should be a consequence of the feed-back of the system
which is able to encode informations on the macroscopic system.
Non linearity and noise are not able to produce localization,
while together they induce the system to localize
in one of its minimal energy configuration positions.
Moreover, the localization positions 
are chosen randomly and they are not fixed in time, 
in the sense that the system can always make a transition
between the two different macroscopic states.
Therefore, it is not possible to predict the result of an observation.

As a final remark, we would like to stress
that our model is based on a genuine quantum dynamics
without more or less phenomenological dissipative terms
in the hamiltonian \cite{Gisin,GRW}.

In our opinion the results presented in this paper
could provide a general framework
for describing the subtle transition
associated with the emergence of the classical world.

\bigskip

It is a pleasure for us to thank Gianfausto Dell'Antonio for fruitful 
discussions and useful suggestions.

\begin{figure}
\begin{center}
\mbox{\psfig{file=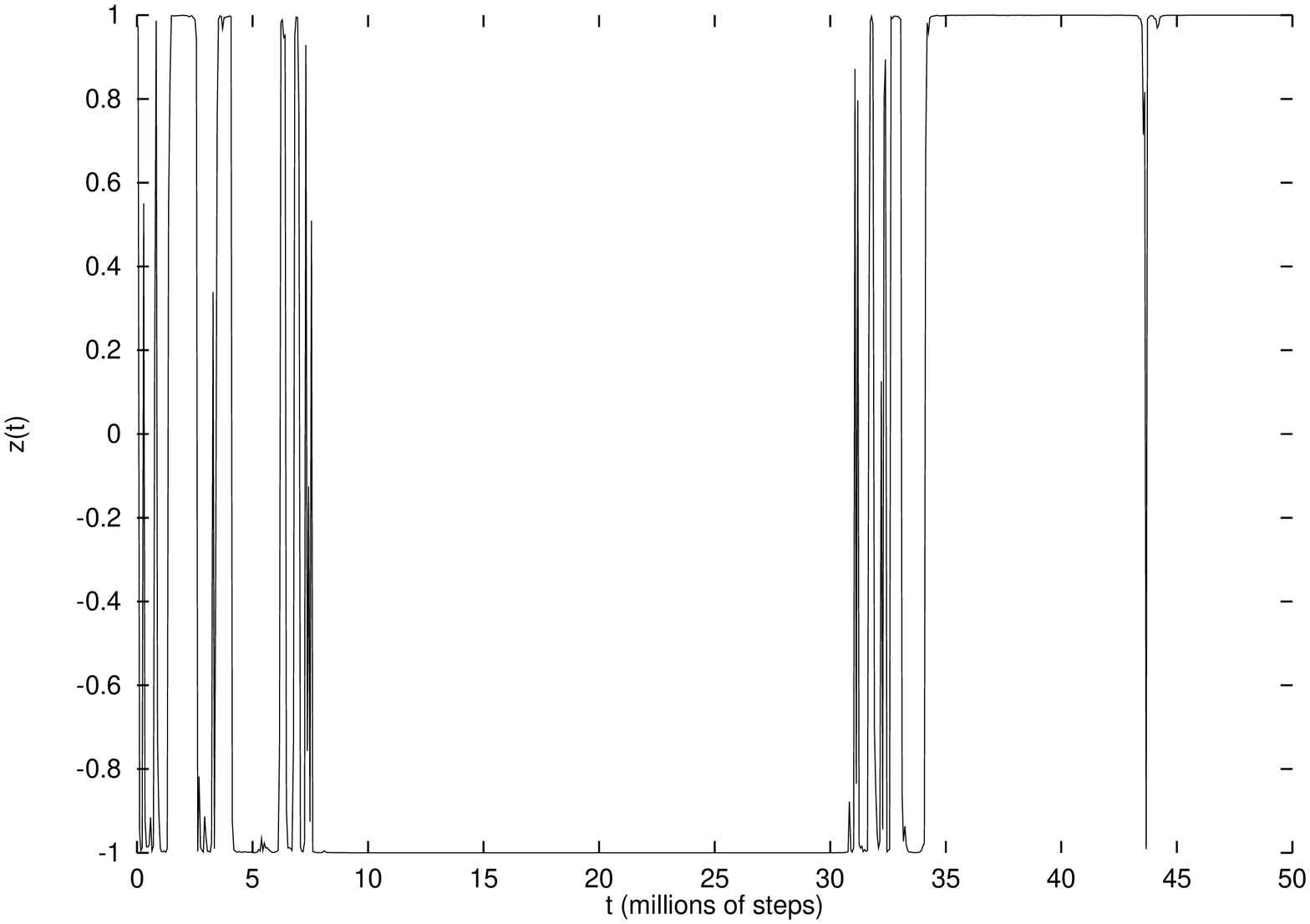,width=12cm,angle=270}}
\end{center}
\caption{A typical realization of $z(t)$ as function of $t$ for $\alpha_0=1$ 
and $\beta=7$. The differential equation [6] is numerically solved
with a time step of $10^{-2}$.}
\end{figure}

\begin{figure}
\begin{center}
\mbox{\psfig{file=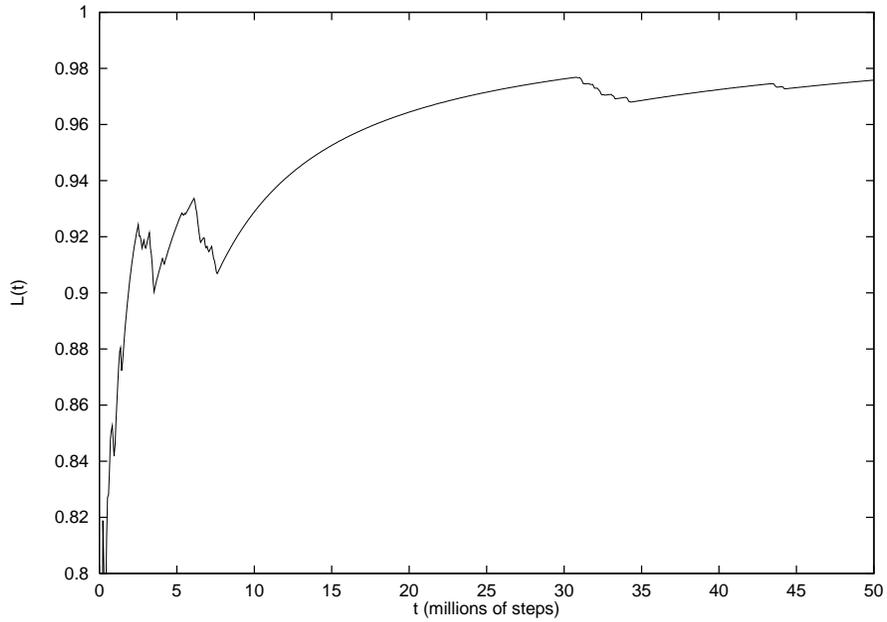,width=12cm,angle=270}}
\end{center}
\caption{$L(t)$ (19) as function of $t$
for the same noise realization of fig. 1, 
with $\alpha_0=1$ and $\beta=7$.}
\end{figure}


\begin{thebibliography}{99}

\bibitem{Gisin}
N. Gisin,
Phys. Rev. Lett {\bf 52} (1984) 1657.

\bibitem{GRW}
G. C. Ghirardi, A. Rimini and T. Weber,
Phys. Rev. D {\bf 34} (1986) 470.

\bibitem{Zurek}
H. Zurek,
Physics Today, (october 1991) 2, and the references therein.

\bibitem{ChKi}
S. Chakravanty and S. Kievelson,
Phys. Rev. Lett. {\bf 50}, (1983) 1811.

\bibitem{ChLe}
S. Chakravanty and A. Legget,
Phys. Rev. Lett. {\bf 52}, (1984) 5.

\bibitem{book}
G. Giulini, E. Joos, C. Kiefer, J. Kupsch, 
I.O. Stamatescu and H.D. Zeh, 
{\it Decoherence and the appearance of 
classical world in quantum theory},  Springer (1996).

\bibitem{GDJH}
F. Grossmann, T. Dittrich, P. Jung and P. H\"anggi,
J. Stat. Phys. {\bf 70}, (1993) 229.

\bibitem{Bell}
J.S. Bell, {\it Speakable and unspeakable in quantum mechanics},
Cambridge University Press, Cambridge (1987).

\bibitem{BBCdAS}
Ph. Blanchard, G. Bolz, M. Cini, G.F. De Angelis and M. Serva,
J. Stat. Phys.  {\bf 75}, (1994) 749.

\bibitem{Berry}
M. Berry, 
in {\it Fundamental Problems in Quantum Theory}, 
eds. D.M. Greenberger and A. Zeilinger, 
Annals of the New York Academy of Sciences, {\bf 755} (1995) 303-317.

\bibitem{CFPSV}
A. Crisanti, M. Falcioni, G. Paladin, M. Serva and A. Vulpiani,
Phys. Rev. E {\bf 50}, (1994) 138.

\bibitem{BDAPS}
Ph. Blanchard, G. Dell'Antonio, M. Pasquini and M. Serva,
in preparation.

\end{thebibliography}
\end{document}